\documentclass[aps,prl,twocolumn,showpacs,preprintnumbers]{revtex4}

\usepackage{graphicx}
\usepackage{dcolumn}
\usepackage{bm}
\usepackage{amsmath}
\usepackage{amssymb}
\usepackage{amsfonts}
\usepackage{float}
\usepackage{hyperref}
\usepackage{dsfont}  
\usepackage{slashed}  
\usepackage{booktabs}
\usepackage{multirow}
\usepackage{subfigure}
\usepackage[sort&compress]{natbib}
\usepackage{xcolor}
\usepackage{color}
\usepackage{colordvi}

\usepackage{graphicx}

\newcommand{\be}{\begin{equation}}  
\newcommand{\ee}{\end{equation}}  
\newcommand{\beq}{\begin{eqnarray}}  
\newcommand{\eeq}{\end{eqnarray}}

\newcommand{\reci}[1]{\frac{1}{#1}}
\newcommand{\Op}{\mathcal{O}}

\renewcommand{\>}{\rangle}
\newcommand{\ms}{\overline{\rm MS}}

\newcommand{\tbl}[1]{Table~\ref{#1}}

\newcommand{\fig}[1]{Fig.~\ref{#1}}

\begin{document}
\title{The nucleon spin and momentum decomposition using lattice QCD simulations}
\author{
  C.~Alexandrou$^{1,2}$,
  M.~Constantinou$^{3}$,
  K.~Hadjiyiannakou$^{1}$,
  K.~Jansen$^{4}$,
  C.~Kallidonis$^{1}$,
  G.~Koutsou$^{1}$,
  A.~Vaquero Avil\'es-Casco$^{5}$,
  C.~Wiese$^{4}$
}
\affiliation{
  $^1$Computation-based Science and Technology Research Center, The Cyprus Institute, 20 Kavafi Str., Nicosia 2121, Cyprus \\
  $^2$Department of Physics, University of Cyprus, P.O. Box 20537, 1678 Nicosia, Cyprus\\
  $^3$Department of Physics, Temple University, 1925 N. 12th Street, Philadelphia, PA 19122-1801, 
USA\\
  $^4$NIC, DESY, Platanenallee 6, D-15738 Zeuthen, Germany\\
  $^5$Department of Physics and Astronomy, University of Utah, Salt Lake City, UT 84112, USA
}


\begin{abstract}
We determine within lattice QCD, the nucleon spin carried by valence
and sea quarks, and gluons. The calculation is performed using an
ensemble of gauge configurations with two degenerate light quarks with
mass fixed to approximately reproduce the physical pion mass. We find
that the total angular momentum  carried by the quarks in the nucleon is
$J_{u+d+s}{=}0.408(61)_{\rm stat.}(48)_{\rm syst.}$ and the gluon
contribution is $J_g {=}0.133(11)_{\rm stat.}(14)_{\rm syst.}$  giving a total of
$J_N{=}0.54(6)_{\rm stat.}(5)_{\rm syst.}$ consistent with the spin sum.
For the quark intrinsic spin contribution we obtain $\frac{1}{2}\Delta
\Sigma_{u+d+s}{=}0.201(17)_{\rm stat.}(5)_{\rm syst.}$. All quantities are given  in the
$\overline{\textrm{MS}}$ scheme at 2~GeV.  The quark and gluon
momentum fractions are also computed and add up to $\langle
x\rangle_{u+d+s}+\langle x\rangle_g{=}0.804(121)_{\rm stat.}(95)_{\rm
  syst.}+0.267(12)_{\rm stat.}(10)_{\rm syst.}{=}1.07(12)_{\rm
  stat.}(10)_{\rm syst.}$ satisfying the momentum sum.
\end{abstract}
\maketitle 
\bibliographystyle{apsrev}


\textit{Introduction:} The distribution of the proton spin amongst its
constituent quarks and gluons has been a long-standing puzzle ever
since the European Muon Collaboration showed in 1987 that only a
fraction of the proton spin is carried by the
quarks~\cite{Ashman:1987hv,Ashman:1989ig}. This was in sharp contrast
to what one expected based on the quark model. This so-called ``proton
spin crisis'' triggered a rich experimental and theoretical activity.
Recent experiments show that only 30\% of the proton spin is
carried by the quarks~\cite{Aidala:2012mv}, while
experiments at RHIC~\cite{Adare:2014hsq,Djawotho:2013pga} on the
determination of the gluon polarization in the proton point to a
non-zero contribution~\cite{deFlorian:2014yva}. A global fit to the most recent experimental data that includes the combined set of inclusive deep-inelastic scattering
data (DIS) from HERA,  
and Drell-Yan data from Tevatron and LHC, led to an improved   determination of the valence quark distributions and the flavor separation of the up- and down-quarks ~\cite{Alekhin:2017kpj}. 
The combined HERA data also provide improved constraints on the  gluon distributions but large uncertainties remain~\cite{Alekhin:2017kpj}.  Obtaining the quark and
gluon contributions to the nucleon spin and momentum fraction within
lattice Quantum Chromodynamics (QCD) provides an independent input that is extremely crucial but the computation is  very challenging. This is
because a complete determination must include, besides the valence  also
sea quark and gluon contributions that exhibit a large
noise-to-signal ratio and are computationally very demanding.  A first
computation of the gluon spin was performed recently via the
evaluation of the gluon helicity in a mixed action approach of overlap
valence quarks on $N_f{=}2{+}1$ domain wall fermions that included an
ensemble with  pion mass 139~MeV~\cite{Yang:2016plb}.
In this work, we evaluate all the contributions to the spin of the
proton as well as the gluon and quark 
momentum fractions~\cite{Alexandrou:2016tuo,Alexandrou:2016mni}.  Such an
investigation has become feasible given the tremendous progress in
simulating QCD on a Euclidean
four-dimensional lattice with quark masses tuned to their physical
values (referred to as the physical point) in combination with new approaches to evaluate sea quark
and gluon contributions that were not possible in the
past~\cite{Alexandrou:2016tuo,Alexandrou:2016ekb,Alexandrou:2016hiy,Constantinou:2017lok}.
This first study of valence and sea quark and gluon contributions  directly  at the physical point allows us to obtain a complete information on the distribution of the nucleon spin and momentum  among its
constituents.


\textit{Computational approach:} We use one gauge ensemble employing
two degenerate ($N_f{=}2$) twisted mass clover-improved
fermions~\cite{Frezzotti:2000nk,Frezzotti:2003ni} with masses that
approximately reproduce the physical pion
mass~\cite{Abdel-Rehim:2015pwa} on a
lattice of $48^3\times96$ and lattice spacing $a{=}0.0938(3)$~fm,
determined from the nucleon mass~\cite{Alexandrou:2017xwd}.  The
strange and charm valence quarks are taken as Osterwalder-Seiler
fermions~\cite{Osterwalder:1977pc,Frezzotti:2004wz}. The  mass of the strange
quark is tuned to
reproduce the $\Omega^-$ mass and the mass of the charm quark is tuned independently to reproduce the mass of $\Lambda_c^+$ as described in detail in Ref.~\cite{Alexandrou:2017xwd}.
 The strange and charm quark masses in lattice units
determined through this matching are $a\mu_s{=}0.0259(3)$ and
$a\mu_c{=}0.3319(15)$, respectively, yielding 
$\mu_c/\mu_s {=} 12.8(2)$. We note that if instead we tune to the  ratio of the kaon (D-meson)  to pion mass  $m_K/m_\pi$ ($m_D/m_\pi$) for the same ensemble we find  $\mu_c/\mu_s {=} 12.3(1)$~\cite{Abdel-Rehim:2015pwa}.  Given that an extrapolation to the  continuum, where the different definitions are expected to be consistent, is not carried out and the errors quoted are only statistical, this level of agreement is very satisfactory.
For the renormalized strange
and charm quark masses we find  $ m_s^R {=}\mu_s/Z_P{=}108.6(2.2)(5.7)(2.6)$~{MeV}
and $ m_c^R {=}\mu_c/Z_P{=}1.39(2)(7)(3)$~{GeV}, where $Z_P$ is the
pseudoscalar renormalization function determined non-perturbatively in
the ${\overline{\rm MS}}$ at
2~GeV~\cite{Alexandrou:2017xwd}. 

\textit{Matrix elements:} We use Ji's sum rule~\cite{Ji:1996ek}, that
provides a gauge invariant decomposition of the nucleon spin as
$$J_{N}{=} \sum_{q{=}u,d,s,c\cdots}\bigg(\reci{2}\Delta\Sigma_q + L_q \bigg)+
J_g,$$
where $\reci{2}\Delta\Sigma_q$ is the contribution from the intrinsic quark spin,
$L_q$ the quark orbital angular momentum and $J_g$ is the gluon total angular momentum.
The quark intrinsic spin $\frac{1}{2}\Delta \Sigma_q$ is obtained from
the first Mellin moment of the polarized parton distribution function
(PDF), which is the nucleon matrix element of the axial-vector
operator.  The total quark angular momentum, $J_q$, can be extracted by computing
the second Mellin moment of the unpolarized nucleon PDF, which is the
nucleon matrix element of the vector one-derivative operator at zero
momentum transfer. These matrix elements in Euclidean space are given
by
\begin{align}
  \langle N(p,s^\prime)|\mathcal{O}^\mu_{A}|N(p,s)\rangle &{=} \bar{u}_N(p,s^\prime) \Bigl[g_A^q\gamma^\mu\gamma_5\Bigr]u_N(p,s),\nonumber\\
  \langle N(p^\prime,s^\prime)| \Op_{V}^{\mu\nu}| N(p,s)\rangle &{=} \bar{u}_N(p^\prime,s^\prime)\Lambda^q_{\mu\nu}(Q^2)u_N(p,s),\nonumber\\
  \Lambda_q^{\mu\nu}(Q^2) {=}  A^q_{20}(Q^2) \gamma^{\{ \mu}P^{\nu\}} &+B^q_{20}(Q^2) \frac{\sigma^{\{ \mu\alpha}q_{\alpha}P^{\nu\}}}{2m}\nonumber\\
  &+C^q_{20}(Q^2)\, \frac{1}{m}Q^{\{\mu}Q^{\nu\}},
\end{align}
with $Q{=}p^\prime{-}p$ the  momentum transfer and $P{=}(p^\prime{+}p)/2$ the
total momentum. The axial-vector operator is $\mathcal{O}^\mu_{A} {=}
\bar{q}\gamma_\mu \gamma_5 q$ and the one-derivative vector operator
$\Op_{V}^{\mu\nu} {=}
\bar{q}\gamma^{\{\mu}\overleftrightarrow{D}^{\nu\}} q$, where the
curly brackets in $\mathcal{O}_{V}$ represent a symmetrization over
pairs of indices and a subtraction of the trace. $\Lambda_q^{\mu\nu}$
is decomposed in terms of three Lorentz invariant generalized form
factors (GFFs) $A^q_{20}(Q^2)$, $B^q_{20}(Q^2)$ and $C^q_{20}(Q^2)$.
A corresponding decomposition can also be made for the nucleon matrix
element of the gluon operator ${\cal {O}}_g^{\mu\nu}$. The quark
(gluon) total angular momentum can be written as 
$J_{q(g)}{=}\reci{2}[A_{20}^{q(g)}(0)+B_{20}^{q(g)}(0)]$, while the
average momentum fraction is determined from
$A^{q(g)}_{20}(0){=}\<x\>_{q(g)}$ and 
$ g_A^q{\equiv}\Delta\Sigma_q$ where $g_A^q$ is the nucleon axial charge.
  While $A_{20}^q(0)$ can be extracted directly
at $Q^2{=}0$, $B^q_{20}(0)$ needs to be extrapolated to $Q^2{=}0$
using the values obtained at finite $Q^2$.

We compute the gluon momentum fraction by considering the $Q^2{=}0$ nucleon
matrix element of the operator ${\cal {O}}_g^{\mu\nu}{=}2 {\rm
  Tr}[G_{\mu\sigma}G_{\nu\sigma}]$, taking the combination
${\cal{O}}_g{\equiv} {\cal {O}}_{44}-\frac{1}{3}{\cal{O}}_{jj}$,
\begin{align}
  \langle N(p,s^\prime)|\mathcal{O}_g|N(p,s)\rangle &{=} \biggr (-4E_N^2-\frac{2}{3}\vec{p}^2\biggl) \langle x\rangle_g,
  \label{gluon}
\end{align}
where we further take the nucleon momentum $\vec{p}{=}0$. 

In lattice QCD the aforementioned nucleon matrix elements are
extracted from a ratio, $R_\Gamma(t_s,t_{\rm ins})$, of a three-point function $G_\Gamma^{\rm 3pt}(t_s,t_{\rm ins})$
 constructed with an  operator $\Gamma$ coupled to a quark divided by
the nucleon two-point
functions $G^{\rm 2pt}(t_s)$,
where $t_{\rm ins}$ is the time slice of the operator insertion
relative to the time slice where a state with the quantum numbers of
the nucleon is created (source).  For sufficiently large time
separations $t_s-t_{\rm ins}$ and $t_{\rm ins}$ the ratio
$R_\Gamma(t_s,t_{\rm ins})$, yields the appropriate nucleon matrix
element. To determine $B_{20}(Q^2)$ we need the nucleon matrix element
for $Q^2\ne 0$, which can be extracted by defining an equivalent ratio
as described in detail in
Refs.~\cite{Alexandrou:2011nr,Alexandrou:2011db,Alexandrou:2010hf}.
An extrapolation of $B_{20}(Q^2)$ is then carried out to obtain
$B_{20}(0)$.  We employ three approaches in order to check that the
time separations $t_s-t_{\rm ins}$ and $t_{\rm ins}$ are sufficiently
large to suppress higher energy states with the same quantum numbers
with the nucleon. These are: i) \emph{Plateau} method. Identify the
range of $t_{\rm ins}$ for which the ratio $R_\Gamma(t_s,t_{\rm ins})$ becomes time-independent and perform a constant
fit; ii) \emph{Summation} method. Summing $R_\Gamma(t_s,t_{\rm ins})$ over $t_{\rm ins}$, to yield
  $\sum_{t_{\rm ins}}R_\Gamma(t_s,t_{\rm ins}){=}R_\Gamma^{\rm sum}(t_s) {=} C +
  t_s\mathcal{M}+\Op\left(e^{-(E_1-E_0)t_s)}\right) + \cdots,$
where $C$ is a constant. The matrix element $\mathcal{M}$ is then
obtained from the slope of a linear fit with respect to $t_s$; iii)
\emph{Two-state fit} method. We perform a simultaneous fit to the
three- and two-point function varying $t_{\rm ins}$ for several values
of $t_s$ include the first excited state in the fit function.  If
excited states are suppressed, the plateau method should yield
consistent values when increasing $t_s$ within a sufficiently large
$t_s$-range.  We require that we observe convergence of the values
extracted from the plateau method and additionally that these values
are compatible with the results extracted from the two-state fit and
the summation method. We take the difference between the plateau and
two-state fit values as a systematic error due to residual excited
states.


The three-point functions for the axial-vector and vector
one-derivative operators entering the ratio $R_\Gamma(t_s,t_{\rm ins})$, 
receive two contributions, one when the operator couples to the
valence up and down quarks (so-called connected) and when it couples
to sea quarks and gluons (disconnected).  The connected contributions
are computed by employing sequential inversion through the
sink~\cite{Martinelli:1988rr}.  Disconnected diagrams are
computationally very demanding, due to the fact that they involve a
closed quark loop and thus a trace over the quark propagator.
A feasible alternative is to employ stochastic
techniques~\cite{Bitar:1989dn} to obtain an estimate of the all-to-all
propagator needed for the evaluation of the closed quark loop.  For
 the up and down quarks, we
utilize \emph{exact deflation}~\cite{Bali:2005fu, Neff:2001zr}, by computing the
$N_{ev}$ lowest eigenmodes of the Dirac matrix to precondition the
conjugate gradient (CG) solver. Taking $N_{ev}{=}500$ yields an
improvement of about twenty times, compared to the standard conjugate
gradient method. 
We also exploit the
properties of the twisted mass action to improve our
computation using the so-called one-end trick~\cite{Michael:2007vn,McNeile:2006bz} that yields an increase in the signal-to-noise ratio~\cite{Alexandrou:2013wca,Abdel-Rehim:2013wlz}.
This also allows the evaluation of
the quark loops for all insertion time-slices, and since the two-point
function is computed for all $t_s$, the disconnected three-point
function is obtained for any combination of $t_s$ and $t_{\rm ins}$
allowing a thorough study of excited states effects.
In addition, an improved approach
is  employed for  $\<x\>_q$
 exploiting the spectral
 decomposition of the Dirac matrix. Within this approach, we use the
 lowest eigenmodes to
construct part of the all-to-all propagator in an \emph{exact} manner.
 This
allows us to invert less stochastic sources for constant variance,
hence $N_r$ is smaller for $\<x\>_q$ in~\tbl{table:statistics}. The
remaining part of the loop is calculated stochastically, with the use
of the one-end trick.

For the heavier  strange and charm quarks, 
the \emph{truncated solver
  method}~\cite{Bali:2009hu} (TSM) performs
well~\cite{Abdel-Rehim:2013wlz,Alexandrou:2013wca}. In the TSM an
appropriately tuned large number of low-precision and a small number
of high-precision stochastic inversions is combined to obtain an
estimate of $G_s(x;x)$. We give the tuned parameters
in~\tbl{table:statistics}. These methods have been recently employed
to compute other nucleon observables using this ensemble
~\cite{Abdel-Rehim:2016won,Alexandrou:2017qyt,Alexandrou:2017hac} as
well as at higher than physical pion
masses~\cite{Abdel-Rehim:2013wlz,Alexandrou:2013wca}.

The three-point function of the gluon operator is purely disconnected. 
To overcome the low signal-to-noise ratio we apply stout smearing to
the gauge links entering the gluonic operator ${\cal
  O}^{\mu\nu}_g$\cite{Morningstar:2003gk}. Use of an analytic link smearing is
essential for performing the perturbative computation of the
renormalization. 
Using smearing and a total of 209400
measurements we obtain the bare matrix element to a few percent accuracy~\cite{Alexandrou:2016ekb}. We note that a non-zero gluon momentum in the quenched approximation was found in Ref.~\cite{Deka:2013zha}.

In~\tbl{table:statistics} we
summarize the statistics used for the calculation for both quark and
gluon observables.
\begin{table}[!h]
  \begin{tabular}{ccc|lcccc}
    \hline\hline
    \multicolumn{3}{c|}{Connected}        &
    \multicolumn{5}{c}{Disconnected}        \\
    $t_s/a$ & $N_{\rm cfg}$ & $N_{\rm src}$ & Observable & $N_{\rm cfg}$  & $N_{\rm src}$ & $N^{HP}_{r}$ & $N^{LP}_{r}$ \\
    \hline
	\multirow{4}{*}{ \begin{tabular}{c}  10,12,14 \\  16 \\  18 \end{tabular} } &
	\multirow{4}{*}{ \begin{tabular}{c}    579    \\ 542 \\ 793 \end{tabular} } &
	\multirow{4}{*}{ \begin{tabular}{c}     16    \\  88 \\  88 \end{tabular} } &
	light,   $g_A$      & 2136 & 100  & 2250 & 0    \\
	&   &    & light,   $\<x\>_q$  & 1219 & 100  & 1000 & 0    \\
    &   &    & strange, $g_A$      & 2153 & 100  & 63   & 1024 \\
	           &   &    & strange, $\<x\>_q$  & 2153 & 100  & 30   & 960  \\
                   &   &    & gluon, $\<x\>_g$ & 2094 &100 & - & -\\
     \hline\hline
  \end{tabular}
  \vspace*{-0.2cm}
  \caption{Statistics used in this calculation. $t_s$ is the sink time
    separation relative to the source which is used for the connected
    three-point functions. $N_{\rm cfg}$ is the number of
    configurations and $N_{\rm src}$ the number of source positions
    per configuration.  $N_r^{HP}$ ($N_r^{LP}$) is the number of high-
    (low-) precision stochastic vectors used for the quark loops.}
  \label{table:statistics}
\end{table}

\textit{Renormalization:} We determine the renormalization functions
for the axial-vector charge and one-derivative vector operators
non-perturbatively, in the RI$^\prime$-MOM scheme. We employ a
momentum source and perform a perturbative subtraction of ${\cal
  {O}}(g^2
a^\infty)$-terms~\cite{Alexandrou:2015sea,Alexandrou:2010me}. This
subtracts the leading cut-off effects yielding only a weak dependence
of the renormalization factors on the renormalization scale $(ap)^2$
for which the $(ap)^2 \rightarrow 0$ limit can be reliably taken. 
Lattice QCD results for both the isovector and isoscalar axial charge
are renormalized non-perturbatively with 
  $Z^{\rm  isovector}_A{=}0.7910(4)(5)$ and $Z^{\rm  isoscalar}_A{=}0.7968(25)(91)$
respectively~\cite{Alexandrou:2015sea,Alexandrou:2017hac}.  The
one-derivative vector operator is non-perturbatively renormalized with $Z_{DV} {=}
1.1251(27)(17)$ in the $\ms$-scheme at
2~GeV~\cite{Alexandrou:2015sea}.
The renormalization of the gluon operator is carried out
perturbatively. Being a flavor singlet operator, it mixes with other
operators and in particular the quark singlet operator.
Due to this mixing, appropriate renormalization conditions require
computation of more than one matrix element. 
We perform the computation in one-loop lattice perturbation theory and use the
action parameters that coincide with the ensemble of this work. To
avoid the introduction
of an intermediate RI-type scheme, we define a convenient
renormalization prescription
that utilizes both dimensional and lattice regularization results (see Ref.~\cite{Alexandrou:2016ekb} for more details).

The physical result of the gluon momentum fraction can be
related to the bare matrix elements  $\langle x \rangle_g^{\rm bare}$ and 
$\langle x \rangle_q^{\rm bare}$ using
%
 $   \langle x \rangle_g = Z_{gg} \langle x \rangle_g^{\rm bare} + Z_{gq}\sum_q \langle x \rangle_q^{\rm bare}$,
where $Z_{gg}$ and $Z_{gq}$ are computed to one-loop. We note that
the mixing coefficient $Z_{gq}$ is a fraction of the statistical errors on
our results. Therefore, for the quark momentum fractions, we
renormalize with the non-perturbatively determined renormalization
factor, neglecting the mixing with the gluon operator. We note that
the perturbative and non-perturbative renormalization functions
$Z_{DV}$ differ by 10\%, which is a much larger effect than the
mixing.

\begin{figure}[ht!]
  \includegraphics[width=\linewidth]{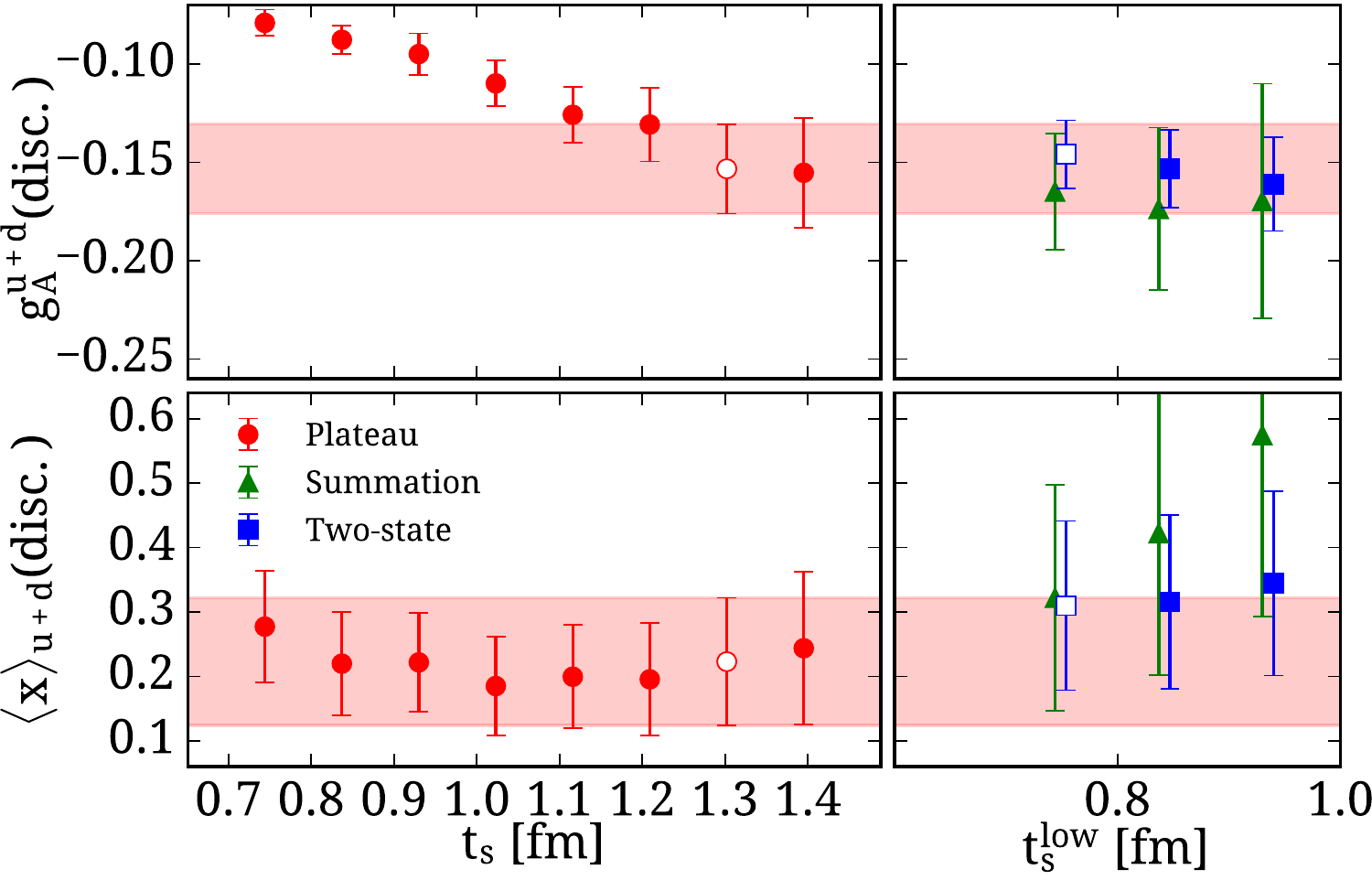}
  \caption{The disconnected sea quark contribution (denoted by disc.) to the
    isoscalar axial charge (upper) and momentum fraction (lower) as a
    function of the sink-source time separation $t_s$ for the plateau
    method (circles) and as a function of the lower time value of
    $t_s$ used in the fits for the summation (green triangles) and
    two-state fit (blue square) methods. The open circle indicates the
    final value and the band its statistical error, while the open
    square is the value taken to determine the systematic error due to
    excited state contamination.}
  \label{fig:gA fits}
\end{figure}
In \fig{fig:gA fits} we show the result of the three analyses carried
out to extract the disconnected contribution to the isoscalar axial
charge $g_A^{u+d}$ and quark momentum fraction $\<x\>_{u+d}$. Taking
the value at $t_s{=}14a{=}1.3$~fm is consistent with the result from the
two-state fit and summation method, for both quantities.  We take the
plateau value at $t_s{=}14a$ as our final result and assign as
systematic error due to excited states the difference between this
value and the mean value determined from the two-state fit. The same
analysis is performed for the strange and charm disconnected
contributions. The analysis for the valence quark contributions at
lower statistics was presented in Ref.~\cite{Abdel-Rehim:2015owa} and
it is followed also here.

\begin{figure}[ht!]
  \includegraphics[width=\linewidth]{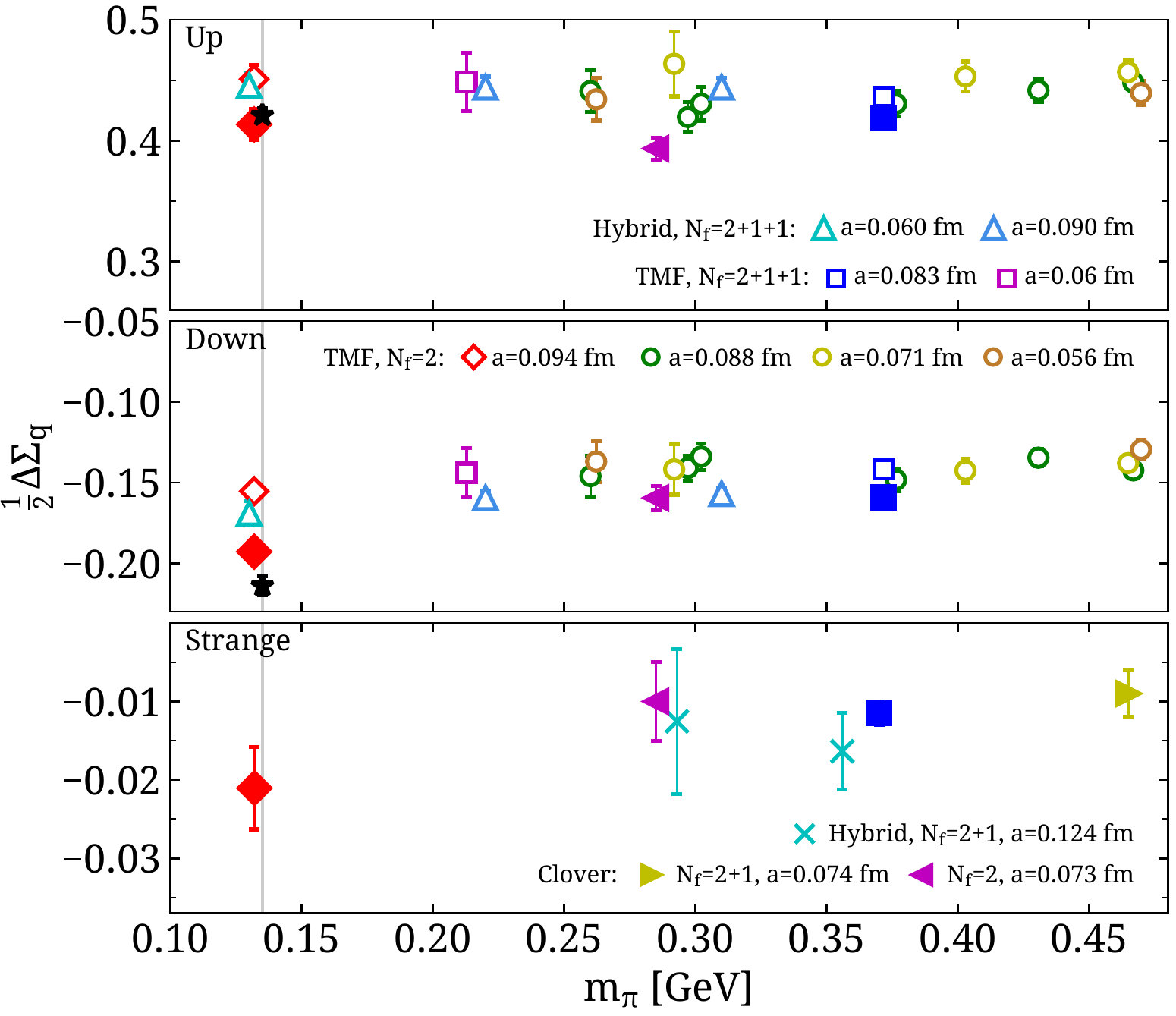}
  \caption{The up (upper), down (center) and strange (lower) quark
    intrinsic spin contributions to the nucleon spin versus the pion
    mass. Open symbols show results with only connected
    contributions while filled symbols denote both connected and disconnected
    contributions using the same ensemble as the one for the connected only.
    Red diamonds are the results of this work.
    Circles are $N_f{=}2$ results, and
    squares are $N_f{=}2{+}1{+}1$~\cite{Alexandrou:2013joa,Abdel-Rehim:2013wlz,Alexandrou:2013wca} by ETMC.
    We compare with lattice QCD results from other ${\cal O}(a)$-improved
    actions from Refs.\cite{QCDSF:2011aa} (filled
    magenta  triangle) by QCDSF, \cite{Engelhardt:2012gd}
    (light blue cross) and~\cite{Chambers:2015bka} by CSSM/QCDSF
    (yellow filled right triangle). We also show results using a hybrid action
    from PNDME~\cite{Bhattacharya:2016zcn} (open blue triangles).
    Experiment is
    denoted by the black asterisks~\cite{Airapetian:2006vy,Blumlein:2010rn}.}
 \label{fig:gAuds}
\end{figure}

 \textit{Results:} In \fig{fig:gAuds} we present our results on
the up, down and strange quark contributions to the nucleon axial
charge that yield the quark intrinsic spin contributions to the
nucleon spin. Since we are using a single ensemble we cannot directly
assess finite volume and lattice spacings effects. However, previous
studies carried out using $N_f{=}2$ and $N_f{=}2{+}1{+}1$ twisted mass fermion (TMF)
ensembles at heavier than physical pion masses for the connected contributions allow us to
assess cut-off and volume effects~\cite{Alexandrou:2011nr,Alexandrou:2013joa}. 
 In \fig{fig:gAuds} 
 we include TMF results for $N_f{=}2$ ensembles   at $m_\pi{\sim} 465$~MeV  one with lattice spacing  $a{=}0.089$~fm  and one with $a{=}0.07$~fm with similar spatial lattice length $L$, as well as,  at $m_\pi{=}260$~MeV, one with $a{=}0.089$~fm  and another with $a{=}0.056$~fm and similar $L$. At both pion masses the results are in complete  agreement as we vary the lattice spacing from $0.089$~fm to $0.056$~fm pointing to cut-off effects smaller than our statistical errors.
For assessing finite volume effects we compare  two $N_f{=}2$ ensembles both with $a{=}0.089$~fm and  $m_\pi {\sim} 300$~MeV, but one with $m_\pi L{=}3.3$ and the other with $m_\pi L{=}4.3$. The values are  completely compatible showing that volume effects are  also within our statistical errors.
To assess possible strange quenching effects 
we compare in Fig.~\ref{fig:gAuds} results  for the connected contributions using  $N_f{=}2$ and $N_f{=}2{+}1{+}1$ TMF ensembles both at $m_\pi{\sim} 375$~MeV
and find very good agreement~\footnote{We find $\frac{1}{2}\Delta \Sigma_u{=} 0.431(11)$ and  $\frac{1}{2}\Delta \Sigma_d{=}  -0.148(7)$ for the $N_f{=}2$ ensemble consistent with $\frac{1}{2}\Delta \Sigma_u{=} 0.436(2)$ and  $\frac{1}{2}\Delta \Sigma_d{=}  -0.142(1)$ for the $N_f{=}2{+}1{+}1$ ensemble.}. The latter is a high statistics analysis yielding very small errors.
 We note, however, that the limited accuracy of the $N_f{=}2$ result would still allow a quenching effect of the order of its statistical error and only an accurate calculation using $N_f{=}2{+}1{+}1$ simulations at the physical point would be able to resolve 
 this completely.
In Fig.~\ref{fig:gAuds}, we also compare recent lattice QCD results on the strange intrinsic spin, $\frac{1}{2}\Delta \Sigma_s$, 
at heavier than physical pion
masses and find agreement among lattice QCD results, indicating that
lattice artifacts are within the current statistical errors. 
We note, in particular, that all lattice QCD results
yield a non-zero and negative strange quark intrinsic spin contribution
$\frac{1}{2}\Delta \Sigma_s$.
We also compute the 
charm axial charge and momentum fraction, at the physical point, and find that both are  consistent with zero.

To determine the total quark angular momentum $J_q$, we need, beyond
$A_{20}^q(0)$, the generalized form factor $B_{20}^q(0)$, which is
extracted from the nucleon matrix element of the vector one-derivative
operator for $Q^2\ne 0$ as described in Ref.~\cite{Alexandrou:2011nr}.
For the isovector case, we find $B_{20}^{u-d}(0)${=}0.313(19), and for
the isoscalar connected contribution
$B_{20}^{u+d,\textrm{conn.}}(0)${=}0.012(20). We observe that the latter
is consistent with zero, as is the disconnected contribution
$B_{20}^{u+d,\textrm{disc.}}(Q^2{=}0.074~{\rm GeV}^2)$.  Similarly, the strange
and charm $B_{20}^{s,c}(Q^2)$ are zero, which implies
$J_{s,c}{=}\frac{1}{2}\langle x\rangle_{s,c}$.  In what follows we will
also take the gluon $B^g_{20}(0)$ to be zero and thus
$J_g{=}\frac{1}{2}\langle x\rangle_g$.

Our final values for the quark total and angular momentum contributions
are given in \tbl{table:results}.
The value of $\langle x \rangle_{u-d}{=}0.194(9)(11)$ is on the upper bound as compared to the recent phenomenological value extracted  in Ref.~\cite{Alekhin:2017kpj}. Determinations of
$\langle x \rangle_{u-d}$ within lattice QCD using simulations with larger than physical pion masses  have yielded larger values, an effect that is partly understood to be due to contribution of excited states to the ground state matrix element~\cite{Bali:2012av}.  We note that our value is in agreement with that determined by RQCD using $N_f{=}2$ clover fermions at pion mass of 151~MeV~\cite{Bali:2014gha} and that  lattice QCD results on
$\langle x \rangle_{u-d}$ and  $J_{u-d}$ for ensembles with larger than physical pion masses including ours are in overall agreement~\cite{Alexandrou:2013joa}.  Results within lattice QCD for the
 individual quark $\langle x \rangle_q$ and $J_q$ contributions are scarce.
 The current computation is the first one using dynamical light quarks with   physical masses. A recent quenched calculation yielded values of $\langle x \rangle_{u,d}$ consistent with ours.

\begin{figure}[ht!]
  \includegraphics[width=0.49\linewidth]{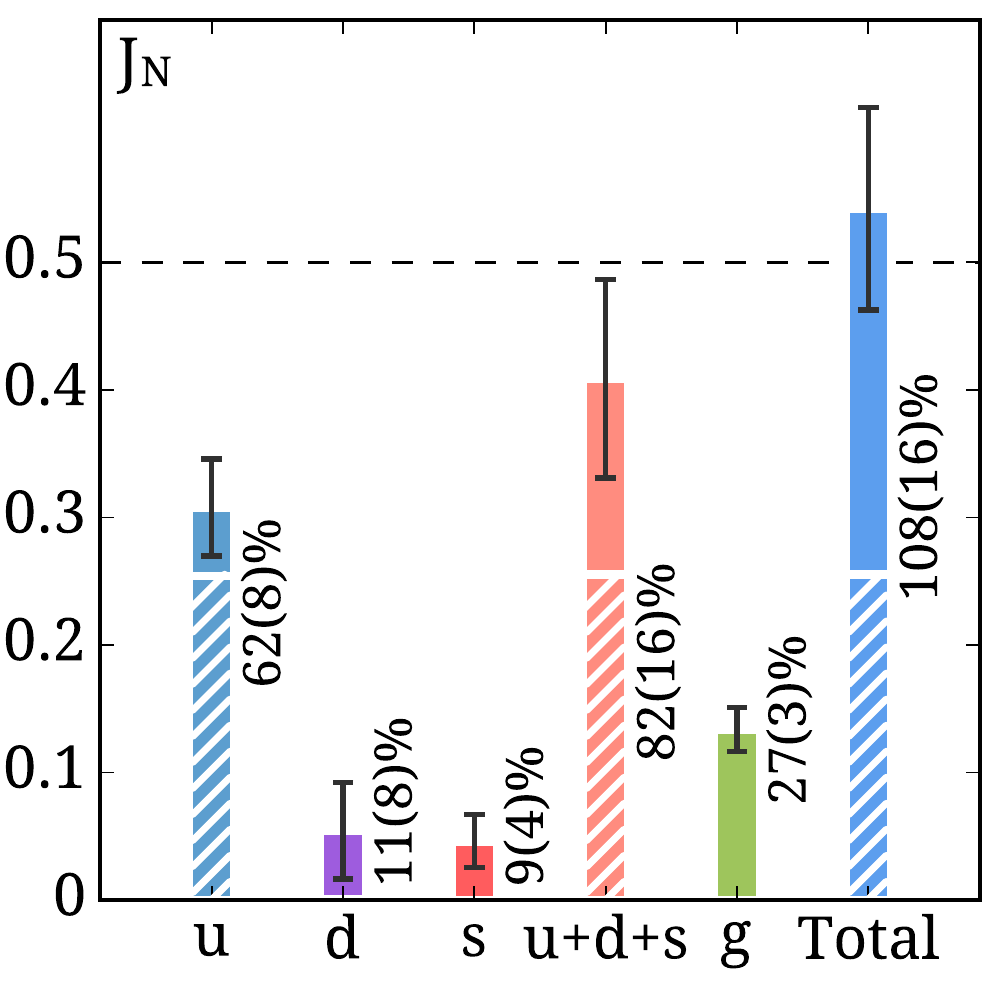}
  \includegraphics[width=0.49\linewidth]{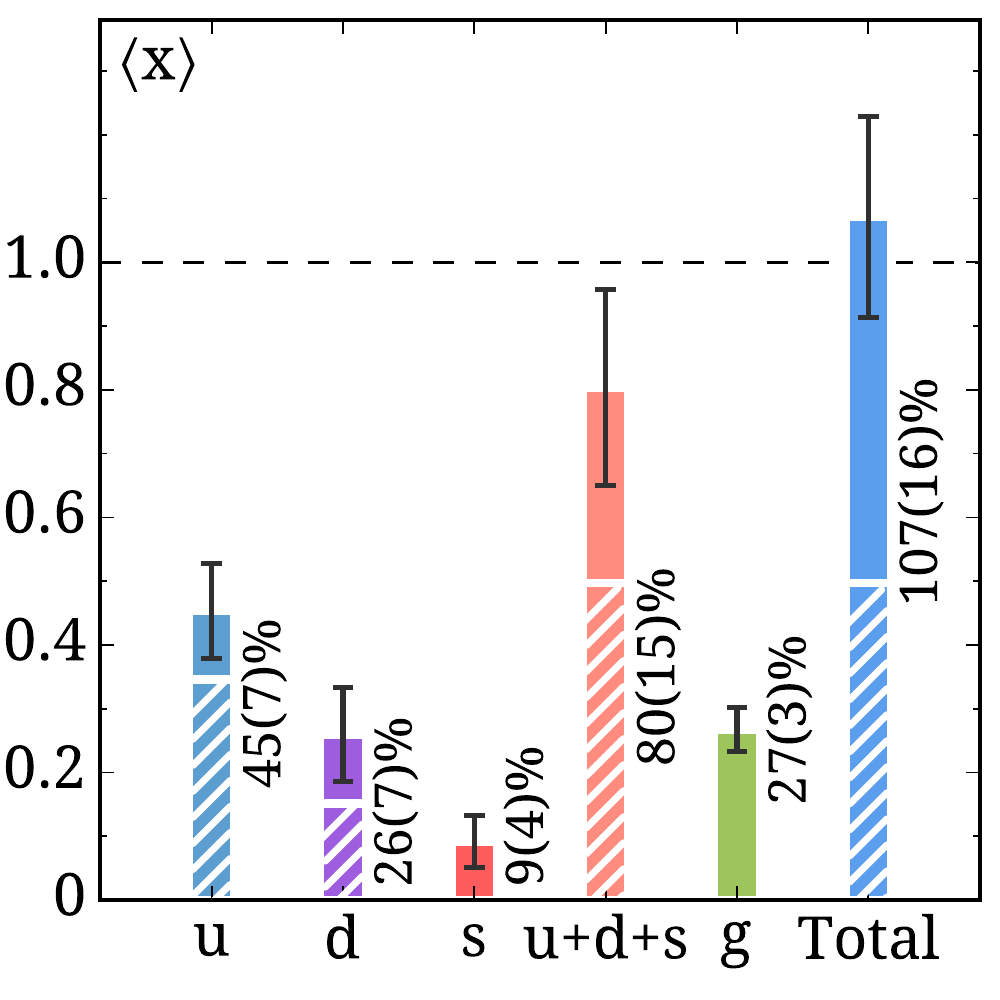}
   \vspace*{-0.4cm}
  \caption{ Left: Nucleon spin decomposition. Right: Nucleon momentum
    decomposition. All quantities are given in the $\overline{\rm
      MS}$-scheme at 2 GeV. The striped segments show valence quark
    contributions (connected) and the solid segments the sea quark and gluon
    contributions (disconnected).}
  \label{fig:pies}
\end{figure}

In \fig{fig:pies} we
show schematically the various contributions to the spin and momentum
fraction. 
Using a different approach to ours, the gluon helicity was recently computed within lattice QCD and found to be
0.251(47)(16)~\cite{Yang:2016plb}. Although we instead compute the gluon total angular momentum and the two approaches  have different systematic uncertainties, we both find non-negligible gluon contributions to the proton spin.

\begin{table}
  \caption{Our results for the intrinsic spin
    ($\frac{1}{2}\Delta\Sigma$), angular  ($L$) and total
    ($J$) momentum contributions to the nucleon spin and to the nucleon
    momentum $\<x\>$, in the $\overline{\rm
      MS}$-scheme at 2 GeV, from up (u), down (d) and strange (s) quarks and
    from gluons (g), as well as the sum of all contributions (tot.),
    where the first error is statistical and the second a systematic
    due to excited states.}
  \label{table:results}
  \begin{tabular}{rr@{.}lr@{.}lr@{.}lr@{.}l}
    \hline\hline
    & \multicolumn{2}{c}{$\frac{1}{2}\Delta\Sigma$} & \multicolumn{2}{c}{$J$} & \multicolumn{2}{c}{$L$}  & \multicolumn{2}{c}{$\<x\>$} \\\hline
          u& 0&415(13)(2)        & 0&308(30)(24)& -0&107(32)(24)        & 0&453(57)(48)\\
          d&-0&193(8)(3)         & 0&054(29)(24)&  0&247(30)(24)        & 0&259(57)(47)\\
          s&-0&021(5)(1)         & 0&046(21)(0) &  0&067(21)(1)         & 0&092(41)(0) \\
          g&\multicolumn{2}{c}{-}& 0&133(11)(14)&  \multicolumn{2}{c}{-}& 0&267(22)(27)\\
       tot.& 0&201(17)(5)        & 0&541(62)(49)&  0&207(64)(45)        & 1&07(12)(10) \\\hline\hline
  \end{tabular}
\label{table:quark spin}
\end{table}

\textit{Conclusions}: In this work we present a calculation of the
quark and gluon contributions to the proton spin, directly at the
physical point.

Having a single ensemble, we can only assess lattice systematic effects due to the
quenching of the strange quark, the finite volume and the lattice
spacing indirectly from other twisted mass ensembles.
A direct evaluation of these systematic errors
is currently not possible and 
will be carried out in the future.
Individual components
are computed for the up, down, strange and charm quarks, including
both connected (valence) and disconnected (sea) quark contributions.
Our final numbers are collected in \tbl{table:results}.  The quark
intrinsic spin from connected and disconnected contributions is
$\frac{1}{2}\Delta\Sigma_{u+d+s}{=}0.299(12)(3)|_{\rm
  conn.}-0.098(12)(4)|_{\rm disc.}{=}0.201(17)(5)$, while the total
quark angular momentum is $J_{u+d+s}{=}0.255(12)(3)|_{\rm
  conn.}+0.153(60)(47)|_{\rm disc.}{=}0.408(61)(48)$. Our result for the
intrinsic quark spin contribution agrees with the upper bound set by a
recent phenomenological analysis of experimental data from
COMPASS~\cite{Adolph:2015saz}, which found $0.13<\reci{2}\Delta\Sigma<
0.18$.
Using the spin sum one would
deduce that $J_g{=}\reci{2}{-}J_q{=}0.092(61)(48)$, which is
consistent with taking $J_g{=}\frac{1}{2}\langle
x\rangle_g{=}0.133(11)(14)$ via the direct evaluation of the gluon
momentum fraction, which suggests that $B_{20}^g(0)$ is indeed small.
Furthermore, we find that the momentum sum is satisfied $
\sum_{q}\langle x \rangle_q+\langle x \rangle_g{=}0.497(12)(5)|_{\rm
  conn.}+0.307(121)(95)|_{\rm disc.}+0.267(12)(10)|_{\rm gluon}{=}1.07(12)(10)$ as is
the spin sum of quarks and gluons giving $
J_N{=}\sum_qJ_q+J_g{=}0.408(61)(48)+0.133(11)(14){=}0.541(62)(49)$ resolving
a long-standing puzzle.

\smallskip
\textit{Acknowledgments:} We thank all members of ETMC for an
enjoyable collaboration and in particular Fernanda Steffens for
fruitful discussions. We acknowledge funding from the European
Union's Horizon 2020 research and innovation program under the Marie
Sklodowska-Curie grant agreement No 642069. M.~C. acknowledges financial support by the National Science Foundation
under Grant No. PHY-1714407. This work used
computational resources from the Swiss National Supercomputing Centre
(CSCS) under project IDs s540, s625 and s702, from the John von
Neumann-Institute for Computing on the Jureca and the BlueGene/Q
Juqueen systems at the research center in J\"ulich from a Gauss
allocation on SuperMUC with ID 44060.

\bibliography{refs}
\end{document}